# Об одной интересной и важной модели в квантовой механике. II. Термодинамическое описание

Юрий Григорьевич Рудой, Енок Олуволе Оладимеджи (мл.)

Российский университет дружбы народов (РУДН)
117198 Россия, Москва, ул. Миклухо-Маклая, 6; e-mail: rudikar@mail.ru, ockjnr@gmail.com

Одной из наиболее интересных моделей в нерелятивистской квантовой механике одной массивной частицы (в одномерном варианте) является модель, введенная Г. Пёшлем и Э. Теллером в 1933 году. Эта модель обладает рядом интересных свойств – в частности, ее предельными случаями являются две наиболее известные модели, изучаемые во всех (в том числе начальных) курсах квантовой механики как для физических, так и для ряда инженерных и технических специальностей. Этими моделями являются модель квазисвободной частицы в «ящике» с непроницаемыми стенками и модель квантового гармонического осциллятора, введенная Ф. Блохом в 1932 году. В предыдущей статье [1] нами подробно рассмотрены динамические свойства модели Пёшля–Теллера и ее предельных случаев, а именно взаимосвязь потенциала, энергетического спектра и механического давления. В настоящей работе дано обобщение полученных результатов на случай отличной от нуля температуры, а именно получены статистическая сумма, термодинамический потенциал и его производные – внутренняя энергия, теплоемкость и давление как в низко-, так и высокотемпературной областях.

*Ключевые слова*: квантовый осциллятор Пёшля–Теллера, статистическая сумма, термодинамические уравнения состояния.

## Введение

Хорошо известно (см., напр., [2, 3]), что все термодинамические, или тепловые, свойства любого ограниченного в пространстве и помещенного в термостат равновесного термодинамического объекта описывается набором *термодинамических уравнений состояния* – термического, калорического и барокалорического. Первое из них связывает давление $P$ с абсолютной температурой $T$ (Кельвина) и объемом $V$ (в одномерном случае – эффективной шириной $L$ конфайнмента), второе – внутреннюю энергию $U$ с $T$ и $V$ (в одномерном случае – соответственно, $L$).

Третье уравнение состояния связывает давление $P$ непосредственно с $U$ и $V$ (или $L$), причем это уравнение особенно широко используется в космологии. Существенно, что по соображениям физической размерности давление $P$ должно выражаться через объемную (в нашем случае – линейную) плотность внутренней энергии $u = U/V$ (или, соответственно, $u = U/L$). Особое место занимает разновидность калорического уравнения состояния, связывающего равновесную энтропию $S$ (Клаузиуса) с температурой $T$ и объемом $V$ (или шириной $L$). Кроме того, большой интерес для термодинамического описания представляют *обобщенные термодинамические восприимчивости* – динамические сжимаемости $\chi_T$ и $\chi_S$ (изотермическая и адиабатическая), а также теплоемкости $C_V$ (или $C_L$), а также $C_P > C_V$ (или $C_L$) (изохорическая и изобарическая).

Существенно, что все указанное разнообразие термодинамических свойств может быть найдено посредством всего лишь одной скалярной величины, а именно термодинамического потенциала $\Psi$, зависящего от любой из пар независимых термодинамических переменных. В нашем случае целесообразно принять в этом качестве пару ($\beta$, $L$), где $L$ – эффективная ширина конфайнмента (см. статью [1]), $\beta = 1/k_B T$ – обратная температура (в энергетических единицах, $k_B = 1{,}38 \cdot 10^{-23}$ Дж/К – постоянная Больцмана). Тогда роль термодинамического потенциала $\Psi(\beta, L)$ играет потенциал Массье–Планка, особенно удобный для нахождения калорических величин – внутренней энергии и теплоемкости.

Хорошо известно (см., напр., [2, 3]), что конкретное вычисление любого термодинамического потенциала (в том числе $\Psi(\beta, L)$) для любого физического объекта всегда основано на знании определенной информации относительно *динамических* свойств этого объекта. Для классического объекта в фазовом пространстве координат $q$ и импульсов $p$ – это знание функции Гамильтона $H(q, p)$ (по существу, энергии объекта). Для квантового объекта – это собственные значения $E_n$, или энергетический спектр оператора Гамильтона ($n = 0, 1, ...,$ – главное квантовое число).

Для простоты будем считать, что $E_n$ – невырожденный чисто дискретный спектр, не имеющий ограничения сверху, но ограниченный снизу энергией основного состояния $E_0$. Существенно, что именно такими свойствами обладает рассматриваемая нами модель квантового осциллятора Пёшля–Теллера (см. формулы (12) и (13) статьи [1]).

Далее, переход от динамики к термодинамике требует введения элементов статистического описания, которое в равновесном случае описывается каноническим распределением Гиббса. В классическом случае функция распределения Гиббса $f(q, p; \beta, L)$ имеет вид

$$f(q, p) = Z^{(-1)}(\beta, L)\exp[-\beta H(q, p)], \qquad (1)$$

в квантовом случае функция распределения Гиббса $f_n(\beta, L)$ имеет вид

$$f_n(\beta; L) \equiv \frac{1}{Z(\beta; L)}\exp[-\beta E_n(L)], \qquad (2)$$

причем выражение (2) определяет вероятность реализации $n$-го квантового состояния с энергией $E_n(L)$ под влиянием теплового движения при отличной от нуля температуре $\beta > 0$. Статистическая сумма $Z(\beta, L)$ определяется из универсального условия нормировки функций (1) или (2)

$$\int f(q,p;\beta,L)dqdp = \sum_n f_n(\beta,L) = 1. \tag{3}$$

Для интересующего нас квантового случая имеем из (3)

$$Z(\beta;L) \equiv \sum_{n=0}^{\infty} \exp[-\beta E_n(L)], \tag{4}$$

так что термодинамический потенциал Массье–Планка дается выражением

$$\Psi(\beta;L) = \ln Z(\beta;L), \tag{5}$$

причем наиболее употребительный в термодинамике потенциал – свободная энергия $F(\beta;L) = (-1/\beta)\Psi(\beta;L)$ очевидным образом связана с $\Psi(\beta;L)$.

Приведем ниже выражения для упомянутых выше уравнений состояния.

*Термическое уравнение состояния* определяется следующим образом:

$$P(\beta;L) = \sum_{n=0}^{\infty}\left[-\frac{\partial E_n(L)}{\partial L}\right]\frac{\exp[-\beta E_n(L)]}{Z(\beta;L)} = \sum_{n=0}^{\infty} P_n(L) f_n(\beta;L), \tag{6}$$

откуда

$$P(\beta;L) = \frac{1}{\beta}\frac{\partial \Psi(\beta;L)}{\partial L}. \tag{7}$$

*Калорическое уравнение состояния* определяется следующим образом:

$$U(\beta;L) = \sum_{n=0}^{\infty} E_n(L) \frac{\exp[-\beta E_n(L)]}{Z(\beta;L)} = \sum_{n=0}^{\infty} E_n(L) f_n(\beta;L), \tag{8}$$

откуда

$$U(\beta;L) = -\frac{\partial \Psi(\beta;L)}{\partial \beta}. \tag{9}$$

Наконец, для *равновесной энтропии* $S(\beta;L)$ получаем следующее выражение, которое может быть записано в виде *соотношения Гиббса–Гельмгольца*:

$$S(\beta;L) = k_B[\Psi(\beta;L) + \beta U(\beta;L)] = k_B[\Psi(\beta;L) - \beta \partial \Psi(\beta;L)/\partial \beta]. \tag{10}$$

На основе термодинамических уравнений состояния (7) и (9) можно определить также введенные выше термодинамические восприимчивости:

$$\chi_{T,S}(\beta,L) = -L[\partial P(\beta,L)/\partial L|_{T,S}], \tag{11}$$

$$C_{L,P}(\beta,L) = -k_B\beta^2[\partial U(\beta,L)/\partial \beta|_{L,P}]. \tag{12}$$

Следующие разделы статьи посвящены вычислению всех выражений (4)–(12) на основе *точного* выражения для энергетического спектра модели Пёшля–Теллера, которое мы здесь для удобства приводим вновь в несколько измененном виде, удобном для дальнейшего вычисления $Z(\beta;L)$:

$$E_n(L,\lambda) = E_0(L,\lambda)\left[\left(\frac{1}{\lambda}\right)n^2 + 2(n+1/2)\right], \quad E_0 = (L,\lambda) = W(L)\lambda(L,V_0). \tag{13}$$

Здесь в соответствии с ранее полученными в [1] формулами имеем

$$\lambda(L,V_0) = 1/2\{1 + [1 + 4v(L,V_0)]^{1/2}\}, \quad v(L,V_0) = \frac{V_0}{W(L)} = \left(\frac{1}{w}\right)V_0 L^2,$$

$$W(L) = wL^{-2}, \quad w = \hbar^2\pi^2/2m. \tag{14}$$

Для дальнейшего существенно, что согласно первой из формул (14) для основного параметра задачи $\lambda(L, V_0)$ в различных предельных случаях модели Пёшля–Теллера имеют место следующие приближенные выражения.

Вблизи *предела свободной частицы* в ящике, когда $V_0 \to 0$, $v(L, 0) \to 0$, что указывает на слабость (или даже полное отсутствие) потенциала $V_0$ внутри ящика. Тогда для вычисления малых поправок на случай учета $V_0 > 0$ имеем

$$\lambda(L,V_0) \approx 1 + v(L,V_0) - 2[v(L,V_0)]^2 + \ldots. \tag{15}$$

Вблизи *предела гармонического осциллятора* $V_0 \to \infty$, $L \to \infty$, когда $v(L, V_0) \to \infty$, а $W(L) \to 0$, но так, что $W(L)v(L, V_0) \sim V_0 L^{-2} \to \mathrm{const}$, имеем

$$1/\lambda(L,V_0) \approx v(L,V_0)^{-1/2} - (1/2)v(L,V_0)^{-1} + \ldots. \tag{16}$$

## Вычисление статистической суммы модели Пёшля–Теллера

К сожалению, для точного спектра (13) вычисление статистической суммы (4) при произвольном значении $\lambda \geq 1$ возможно только приближенно, поэтому, следуя методике [2, 3], мы рассмотрим порознь *низко- и высокотемпературные области*.

Заметим, что такой подход характерен практически для всех известных квантово-статистических моделей, в том числе и для свободной частицы в ящике при $\lambda = 1$, когда из (13) прямо следует

$$E_0(L,1) = W(L), \quad E_n(L) = E_0(L,1)(n+1)^2, \quad n = 0,1,\ldots, \tag{17}$$

так что сложный спектр (13) «сворачивается» в полный квадрат по *n*.

Тогда статистическая сумма (4) принимает вид т.н. *гауссовской суммы*

$$Z^{\mathrm{СЧ}}(\beta,L) = \sum_{n=0}^{\infty}\exp[-\beta W(L)(n+1)^2] = \sum_{n=1}^{\infty}\exp[-\beta W(L)n^2], \tag{18}$$

которая выражается через одну из Θ-функций Якоби, не имеющей аналитического выражения через какие-либо элементарные функции, но все же допускающей низко- и высокотемпературные разложения (подробнее см. [4]). Заметим, что сдвиг целочисленного индекса суммирования (представляющего главное квантовое число *n*) на единицу устраняет как *линейное* по *n*, так и *постоянное* слагаемое в показателе экспоненты (18).

Случаю гармонического осциллятора Блоха отвечает противоположный предел $\lambda \to \infty$, причем в точке $1/\lambda = 0$ из спектра (3.13) устраняется – в отличие от предела свободной частицы – как раз *квадратичное* слагаемое, так что

$$E_0(\infty,\infty) = (1/2)\hbar\omega_0, \quad E_n(\infty) = 2E_0(\infty,\infty)\left(n+\frac{1}{2}\right). \tag{19}$$

Тогда статистическая сумма (3.4) принимает вид бесконечной геометрической прогрессии со знаменателем $\exp[-\beta\hbar\omega_0]$ (который всегда $< 1$):

$$Z^{\text{ГО}}(\beta) = \sum_{n=0}^{\infty} \exp[-\beta\hbar\omega_0(n+1/2)]. \tag{20}$$

Только этот случай допускает *точное* суммирование и потому является в определенном смысле «эталонным» в квантово-статистической проблематике

$$Z^{\text{ГО}}(\beta) = \exp(-1/2\beta\hbar\omega_0)/[1-\exp(-\beta\hbar\omega_0)] = 1/2\operatorname{ch}1/2\beta\hbar\omega_0. \tag{21}$$

В общем случае модели квантового осциллятора Пёшля–Теллера для статистической суммы (4) со спектром (13) имеем

$$Z^{\text{ПТ}}(\beta, L, \lambda) = \exp[-\beta E_0(L,\lambda)]\zeta(\beta, L), \tag{22}$$

где «малая» статистическая сумма $\zeta(\beta, L)$ имеет вид[1]

$$\zeta(\beta, L, \rho\rho\lambda) = \sum_{n=0}^{\infty} \exp\{-\beta E_0(L,\lambda)\}[(1/\lambda)n^2 + 2n]. \tag{23}$$

Тогда введенный в (3.5) термодинамический потенциал Масье–Планка принимает вид

$$\Psi(\beta; L, \lambda) = -\beta E_0(L,\lambda) + \psi(\beta; L, \lambda), \quad \psi(\beta; L, \lambda) = \ln \zeta(\beta; L, \lambda). \tag{24}$$

С целью перехода к приближенным вычислениям для величин $\zeta(\beta; L, \lambda)$, $\psi(\beta; L, \lambda)$, $\Psi(\beta; L, \lambda)$ – и, в конечном счете, уравнений состояния (7), (9) и (10), введем, следуя стандартным процедурам (см, например, [2–4]), безразмерную (обратную) температуру, нормированную на *характеристическую температуру* $\beta_0(\lambda)$:

$$\widetilde{\beta}(L,\lambda) \equiv \beta/\beta_0(L,\lambda), \quad \beta_0(L,\lambda) = 1/k_B T_0(L,\lambda), \quad T_0(L,\lambda) = E_0(L,\lambda)/k_B. \tag{25}$$

Далее можно условно разделить весь интервал температур на две области – соответственно, *низкотемпературную* (по существу чисто квантовую)

$$\widetilde{\beta}(L,\lambda) \gg 1, \quad \beta \gg \beta_0(L,\lambda), \quad T \ll T_0(L,\lambda), \tag{26}$$

---

[1] Для упрощения записи здесь и далее мы опускаем явное указание аргументов $L$ и $V_0$ при записи параметра $\lambda$; в случае необходимости следует применять выражения (15) и (16).

и *высокотемпературную* (по существу квазиклассическую)

$$\widetilde{\beta}(L,\lambda) \ll 1, \quad \beta \ll \beta_0(L,\lambda), \quad T \gg T_0(L,\lambda); \tag{27}$$

в соответствии с этим аппроксимируем далее «малую» статистическую сумму $\zeta(\beta, L, \lambda)$ из (22) адекватным образом в областях (25) и (26).

## Низкотемпературная область для модели Пёшля–Теллера

В области низких температур (25), когда значение $\widetilde{\beta}(\beta, L, \lambda)$ достаточно велико, бесконечную сумму для $\zeta(\beta, L)$ из (22) можно аппроксимировать несколькими первыми слагаемыми. Разумеется, это возможно лишь при условии, что строго дискретные уровни энергии $E_n$ монотонно возрастают с номером уровня $n$, что очевидным образом выполняется для нашей модели:

$$\zeta(\beta, L, \lambda) = \sum_{n=0}^{\infty} \exp\{-\widetilde{\beta}(L,\lambda)[E_n(L,\lambda) - E_0(L,\lambda)]/E_0(L,\lambda)\} =$$

$$= \sum_{n=0}^{\infty} \exp\{-\widetilde{\beta}(L,\lambda)[(1/\lambda)n^2 + 2n]\}. \tag{28}$$

Ограничиваясь для простоты первыми тремя слагаемыми, имеем:

$$\zeta(\beta, L, \lambda) \approx 1 + \exp\{-\widetilde{\beta}(L,\lambda)[(1/\lambda) + 2]\} + \exp\{-4\widetilde{\beta}(L,\lambda)[(1/\lambda) + 1]\} + \ldots; \tag{29}$$

учитывая затем приближенное равенство $\ln(1 + x) \approx x$ при $x \to 0$, получаем для потенциала Масье–Планка (23) в низшем приближении:

$$\Psi(\beta; L, \lambda) = -\widetilde{\beta}(L,\lambda) + \exp\{-\widetilde{\beta}(L,\lambda)[(1/\lambda) + 2]\} + \ldots. \tag{30}$$

Легко видеть, что по сравнению с первым (основным) слагаемым $[-\widetilde{\beta}(L, \lambda)]$ все остальные слагаемые в (29) имеют экспоненциальную малость. Это обусловлено строго дискретной структурой спектра энергий и наличием в нем «энергетической щели» между основным и первым возбужденным состояниями осциллятора в модели Пёшля–Теллера.

Ясно, что это же свойство малости будет характерно также и для всех термодинамических величин (7)–(12), связанных с потенциалом $\Psi(\beta; L, \lambda)$ дифференциальными или алгебраическими соотношениями. Например, учитывая следующее из (24) равенство $\partial\widetilde{\beta}/\partial\beta = 1/\beta_0 = k_B T_0 = E_0$, имеем для внутренней энергии в модели Пёшля–Теллера при низких температурах:

$$U^{\text{ПТ}}(\beta; L, \lambda) = E_0(L,\lambda)\{1 + [(1/\lambda) + 2]\exp[-\widetilde{\beta}(L,\lambda)(1/\lambda) + 2)]\}. \tag{31}$$

Видно, что выражение (31) правильно воспроизводит известные выражения (см., например, [5]) для внутренней энергии свободной частицы в ящике

$$U^{\text{СЧ}}(\beta; L, \lambda) = E_0(L,1)\{1 + 3\exp[-3T_0(L,1)/T]\}, \quad E_0(L,1) = W(L), \tag{32}$$

а также для гармонического осциллятора Блоха ($L \to \infty, \lambda \to \infty$)

$$U^{\text{ГО}}(\beta) = E_0^{\text{ГО}}\{1 + 2\exp[-2T_0^{\text{ГО}}/T]\}, \quad E_0^{\text{ГО}} = (1/2)\hbar\omega_0. \qquad (33)$$

Следующей важной калорической величиной является теплоемкость; дифференцируя согласно (12) выражение (30) по $\beta$, получаем

$$C_L(\beta, L) = -k_B[\widetilde{\beta}(L, \lambda)]^2[(1/\lambda) + 2]^2 \exp\{-\widetilde{\beta}(L, \lambda)[(1/\lambda) + 2]\}, \qquad (34)$$

откуда нетрудно получить хорошо известные частные случаи для СЧ и ГО. Обратим внимание, что теплоемкость (33) обращается в нуль в пределе низких температур ($T \to 0$, $\widetilde{\beta} \to \infty$) в полном соответствии с третьим началом термодинамики, поскольку множитель $\widetilde{\beta}$ («большой» в пределе $\widetilde{\beta} \to \infty$) «подавляется» экспоненциально малым множителем $\exp[-\widetilde{\beta}(L, \lambda)(1/\lambda) + 2)]$.

Рассмотрим еще одну калорическую величину — равновесную энтропию Клаузиуса, подставив в формулу (10) потенциал (30) и внутреннюю энергию (31):

$$S(\beta; L, \lambda)/k_B = \Psi(\beta; L, \lambda) + \beta U(\beta; L, \lambda) \approx -\widetilde{\beta}(L, \lambda) + \exp\{-\widetilde{\beta}(L, \lambda)[(1/\lambda) + 2]\} +$$
$$+ \widetilde{\beta}(L, \lambda) + \widetilde{\beta}(L, \lambda)[(1/\lambda) + 2]\exp\{-\widetilde{\beta}(L, \lambda)[(1/\lambda) + 2]\} \to 0$$
$$\text{при } \widetilde{\beta} \to \infty, T \to 0. \qquad (35)$$

Обратим внимание на точную компенсацию «больших» слагаемых вида $\widetilde{\beta}(L, \lambda)$ в выражении для безразмерной (в единицах $k_B$) энтропии, что обеспечивает ее экспоненциально быстрое обращение в нуль при нуле температуры. Заметим, что нулевое значение $S$ при $T = 0$ находится в полном соответствии с невырожденностью спектра (в том числе энергии основного состояния) модели Пёшля–Теллера.

Наконец, рассмотрим термическую величину — давление $P(\beta; L, \lambda)$, определенное равенством (7); здесь мы ограничимся полуколичественным анализом ввиду очевидно малого влияния экспоненциально малых слагаемых. Используя приближенное выражение (30) для $\Psi(\beta; L, \lambda)$, получаем для давления (с точностью до экспоненциально малых слагаемых)

$$P(\beta; L, \lambda) = (1/\beta)\partial/\partial L\{-\widetilde{\beta}(L, \lambda) + O[\exp-\widetilde{\beta}(L, \lambda)((1/\lambda) + 2)]\} \approx (1/\beta)\partial/\partial L[-\widetilde{\beta}(L, \lambda)] =$$
$$= (1/\beta)\partial/\partial L[-\beta/\beta_0(L, \lambda)] = -\partial/\partial L[1/\beta_0(L, \lambda)] = -\partial/\partial L[E_0(L, \lambda)] = P_0(L, \lambda). \qquad (36)$$

Поясним этот результат: поскольку, как уже отмечалось, в пределе $T \to 0$ все состояния, лежащие выше основного, возбуждаются с экспоненциально. Как уже отмечалось, для калорических величин — энтропии и теплоемкости это приводит к нулевым значениям, тогда как для термических — энергии и давления — мы получаем для них чисто динамические (вообще говоря, отличные от нуля) значения — энергию основного состояния $E_0(L, \lambda)$ и соответствующее ему давление $P_0(L, \lambda)$.

## Высокотемпературная область для модели Пёшля–Теллера

Рассмотрим теперь область высоких температур, для которой выполняются условия (27) и значения величины $\widetilde{\beta}(L,\lambda) = \beta E_0(L,\lambda)$ малы. При этом суммируемая в (23) функция $\rho(n) = \exp\{-\widetilde{\beta}(L,\lambda)[(1/\lambda)n^2 + 2n]\}$ удовлетворяет всем требованиям метода Эйлера–Маклорена (см., например, [3]). Это позволяет перейти от суммы по дискретному индексу $n$ к интегралу по $x$, так что $\zeta(\beta,L) = \sum_{n=0}^{\infty}\rho(n)$ переходит в $\int_{\infty}^{0}\rho(x)dx + (1/2)[\rho(0) + \rho(\infty)]$ +другие внеинтегральные члены, содержащие все производные нечётных порядков функции $\rho(x)$ на концах интервала интегрирования при $x = 0$ и $x \to \infty$. Нетрудно видеть, что все эти внеинтегральные слагаемые дают степенные поправки возрастающего порядка малости по параметру $\beta \to 0$.

Ограничиваясь вкладом только интегрального члена и применяя интегральную формулу Пуассона, находим (временно применяя сокращённые обозначения):

$$\zeta(\widetilde{\beta};\lambda) = \Gamma(1)(2\widetilde{\beta}/\lambda)^{-1/2}\exp[(1/2)\widetilde{\beta}\lambda]D_{-1}[(2\widetilde{\beta}\lambda)^{1/2}], \qquad (37)$$

где $\Gamma(1) = 1$ – значение гамма-функции, $D_{-1}$ – функция параболического цилиндра, или функция Вебера, с отрицательным индексом (–1). Последняя связана с функцией Эрмита $H_{-1}$ соотношением, содержащим т.н. функцию ошибок, или интеграл вероятности $\Phi(y) = (2/\sqrt{\pi})\int_0^y dt\exp(-t^2)$:

$$D_{-1}(y) = \sqrt{2}\exp[-y^2/4]H_{-1}(y/\sqrt{2}),$$
$$H_{-1}(y/\sqrt{2}) = (\sqrt{\pi}/2)\exp[-y^2/2]\{1 - \Phi(y/\sqrt{2})\};$$

в дальнейшем будут использованы асимптотические разложения для $\Phi(y)$:

$$\Phi(y) \approx (2\sqrt{\pi})y \quad (y \ll 1), \quad 1 - \Phi(y) \approx (1/\sqrt{\pi})(1/y)\exp(-y^2) \quad (y \gg 1). \qquad (38)$$

В результате «малая» статистическая сумма (37) принимает вид:

$$\zeta(\widetilde{\beta};L,\lambda) = (\sqrt{\pi}/2)(\lambda/\widetilde{\beta})^{1/2}\exp(\lambda\widetilde{\beta})\{1 - \Phi[(\lambda\widetilde{\beta})^{1/2}]\}, \qquad (39)$$

причём уже в этом выражении можно усмотреть некоторые простые, но важные для дальнейшего анализа факты. Прежде всего, множитель $(\lambda/\widetilde{\beta})^{1/2} = [\lambda/\beta W(L)\lambda]^{1/2} = [\beta W(L)]^{-1/2}$ вообще не содержит параметра $\lambda$, тогда как для входящих в (39) функций аргумент $(\lambda\widetilde{\beta}) = \lambda^2[\beta W(L)]$ пропорционален $\lambda^2$.

Эти факты важны для получения выражений в предельных случаях модели ПТ ($\lambda = 1$ для СЧ в ящике и $1/\lambda = 0$ для ГО Блоха), а также поправок к ним за счёт конечности величин $V_0 > 0$ и $L > 0$ соответственно. Согласно (5), полный термодинамический потенциал Масье–Планка (24) для общего случая модели Пёшля–Теллера с произвольным значением $\lambda \geq 1$ имеем

$$\Psi(\widetilde{\beta};L,\lambda) = -\widetilde{\beta} + \ln\zeta(\beta;\lambda) \approx$$
$$\approx A + \beta V_0 - (1/2)\ln[\beta W(L)] + \ln\{1 - \Phi[\lambda(\beta W(L))^{1/2}]\}, \qquad (40)$$

где постоянная $A = \ln(\sqrt{\pi}/2)$ и учтено, что $\widetilde{\beta} = \beta/\beta_0$, где $1/\beta_0 = E_0(\lambda) = W(L)\lambda$. Заметим, что с учетом точного определения (11) (см. статью [1]) $\lambda(\lambda - 1) = \nu = V_0/W(L)$ можно исключить $\lambda$ из слагаемого $\widetilde{\beta}(\lambda - 1)$ в (40), так что $\beta W(L)(V_0/W(L)) = \beta V_0$; таким образом, параметр $\lambda$, отличающий модель Пёшля–Теллера от моделей СЧ и ГО, входит лишь в одно слагаемое термодинамического потенциала.

Поскольку в этом разделе рассматривается только высокотемпературная область, значение приведенной обратной температуры $\widetilde{\beta}$ всюду является малым (но все же отличным от нуля), тогда как параметр $\lambda$ в пределе ГО может становиться большим (строго говоря, бесконечным). Поэтому предельные случаи (квази)свободной частицы ящике с конечным $\lambda$ и гармонического осциллятора с $\lambda \to \infty$ следует рассматривать по отдельности.

**Предел квазисвободной частицы в ящике (конечное значение $\lambda$).** В этом случае

$$\Psi(L, \widetilde{\beta}; \lambda) = -\widetilde{\beta} + \ln \zeta(\beta; \lambda) \approx$$
$$\approx A + \beta V_0 - (1/2)\ln[\beta W(L)] - (2/\sqrt{\pi})\lambda(\beta W(L))^{1/2}. \tag{41}$$

Для средней энергии согласно (9) находим

$$U(\beta; L, \lambda) = -\partial \Psi(\beta; L, \lambda)/\partial \beta = -V_0 + (1/2)(1/\beta)\{1 + (2/\sqrt{\pi})\lambda[W(L)\beta]^{1/2}\}. \tag{42}$$

Для теплоемкости (при постоянном значении ширины $L$) имеем тогда

$$C_L(\beta; L, \lambda) = -k_B \beta^2 [\partial U(\beta; L, \lambda)/\partial \beta] = k_B (1/2)\{1 + (1/\sqrt{\pi})\lambda[W(L)\beta]^{1/2}\}. \tag{43}$$

$\theta(x) = (1/\sqrt{x})\theta(1/x)$ Заметим, что при *конечных* значениях $\lambda \geq 1$ теплоемкость (43) проявляет тенденцию к уменьшению за счет уменьшения внутренней энергии (42); этот результат находится в качественном соответствии с работой [5].

Представляет интерес сравнение результата (43) с хорошо известным (см., напр., [4]) для предельного случая свободной частицы в ящике при $\lambda = 1$, или, что то же, $V_0 = 0$. Нетрудно показать, что статистическая сумма в этом случае выражается через т.н. гауссовскую сумму, выражающуюся через одну из *функций Якоби*:

$$\Theta(x) = \sum_{n=-\infty}^{+\infty} \{\exp[-\pi n^2 x]\}, \quad \Theta(x) \to 1 \text{ при } x \to \infty, \tag{44}$$

поскольку все члены суммы равны нулю кроме члена с $n = 0$; очевидно, в нашем случае $x = \beta W(L)/\pi$, причем в высокотемпературном пределе $x \to 0$. Для перехода к этому случаю удобно использовать полезное свойство $\theta$-функции, а именно функциональное уравнение вида            . Имеем тогда

$$Z(\beta; L,1) = (1/2)[\theta(\beta W(L)/\pi) - 1] = (1/2)\{(\beta W(L)/\pi)^{-1/2} \theta[\pi/\beta W(L)] - 1\}; \tag{45}$$

аппроксимируя далее $\theta(1/x) = 1$ в высокотемпературном пределе $x \to 0$, находим для $Z(\beta; L, 1)$ приближенное выражение

$$Z(\beta; L,1) \approx (1/2)(\beta W(L)/\pi)^{-1/2}\{1 - (\beta W(L)/\pi)^{1/2}\},$$

откуда легко получить в том же приближении термодинамический потенциал

$$\Psi(\beta;L,1) = \ln Z(\beta;L,1) = A - (1/2)\ln[\beta W(L)/\pi] - (\beta W(L)/\pi)^{1/2},$$
$$A = \ln(\sqrt{\pi}/2) \tag{46}$$

практически полностью совпадающий с полученным нами выше в (41).

**Предел гармонического осциллятора ($\lambda \to \infty$).** Для получения этого предела необходимо вернуться к общему выражению для термодинамического потенциала (40) и рассмотреть предел для последнего слагаемого в случае, когда $\beta$ мало, но конечно, тогда как $\lambda \to \infty$. Учтем асимптотику функции ошибок $\Phi(y)$ при $y \to \infty$, где в этом случае $y = \lambda(\beta W(L)^{1/2}$:

$$1 - \Phi(y) \approx (1/\sqrt{\pi})(1/y)\exp(-y^2), \quad \ln[1 - \Phi(y)] \approx -(1/2)\ln\pi - \ln y - y^2. \tag{47}$$

Учтем далее, что согласно результатам главы 1 (см. формулу (14)) переход к пределу гармонического осциллятора $\lambda \to \infty$ обязательно сопровождается одновременным переходом к пределу $L \to \infty$, $W(L) \to 0$, причем существует конечный предел[2] величины $\lambda W(L)$, равный $(1/2)\hbar\omega_0$. Тогда величина $y = [(1/2)\beta\hbar\omega_0]^{1/2}$ становится малой в высокотемпературном пределе, так что в (47) можно пренебречь $y^2$ по сравнению с $\ln y$ и получить окончательно для гармонического осциллятора в этом пределе выражение,

$$\Psi(\beta;\omega_0) \approx \widetilde{A} + (1/2)\beta\hbar\omega_0 - \ln\{(1/2)\beta\hbar\omega_0\}, \tag{48}$$

которое правильно воспроизводит термодинамику ГО при высоких $T \to \infty$.

**Термическое уравнение состояния для квазисвободной частицы в ящике.** Завершим анализ термодинамических следствий модели Пёшля–Теллера построением термического уравнения состояния[3], связывающего давления с эффективной шириной $L$ конфайнмента и (обратной) температурой $\beta$. Дифференцируя термодинамический потенциал (41) по $L$ в соответствии с определением (7) и используя формулу (11) из статьи [1], находим:

$$P(\beta;L,\lambda) = (1/\beta)\partial\Psi(\beta;L,\lambda)/\partial L = (1/\beta)(1/L)\{1 + (2/\sqrt{\pi})\lambda[W(L)\beta]^{1/2}\}. \tag{49}$$

Сравнивая выражения (49) для давления и (42) для внутренней энергии, нетрудно установить наличие простой связи между этими величинами, иногда называемое *барокалорическим* (или бароэнергетическим) уравнением

$$P(\beta;L,\lambda) = (2/L)[U(\beta;L,\lambda) + V_0], \quad P(\beta;L,\lambda) = 2u(\beta;L,\lambda),$$
$$u(\beta;L,\lambda = (1/L)[U(\beta;L,\lambda) + V_0]. \tag{50}$$

---

[2] Аналогичную процедуру предельного перехода следует провести и для первого слагаемого вида $\widetilde{\beta}(\lambda - 1) \approx (1/2)\beta\hbar\omega_0$.

[3] Как уже неоднократно отмечалось, подобное уравнение состояние не имеет смысла в пределе гармонического осциллятора ввиду необходимости перехода к пределу $L \to \infty$, при котором понятие давления не может быть корректно определено.

Таким образом, в высокотемпературной области модель Пёшля–Теллера сохраняет *линейную связь* между давлением и плотностью (здесь – одномерной в расчете на единицу длины $L$) внутренней энергии (сдвинутой на несущественную постоянную $V_0$). Как известно (см., например, [2–4]), общий вид подобной связи имеет вид $P = (k/f)u$, где объемная плотность внутренней энергии $u = U/L^f$, $f$ – число степеней свободы (в нашем случае $f = 1$), $k$ – показатель однородности в законе дисперсии $E_{кин}(p) \sim p^k$ кинетической энергии частицы $k$ (в нашем случае для нерелятивистской частицы $k = 2$, для релятивистской частицы $k = 1$).

В завершение данного раздела рассмотрим по отдельности вклады, которые вносят в давление (а согласно (50) и в плотность внутренней энергии) порознь *квантовые эффекты* (пропорциональные постоянной Планка $\hbar$) и эффекты, обусловленные потенциалом Пёшля–Теллера (пропорциональные амплитуде $V_0$ этого потенциала).

Действительно, величина $(1/I)(1/L)$ представляет собой выражение для давления, получаемое в рамках чисто *классических* представлений, согласно которым $P = nk_BT$; этому выражению соответствует первое слагаемое (=1) в фигурных скобках выражения (49).

Далее, для свободной квантовой частицы – с учетом квантования уровней энергии, но в отсутствие потенциала Пёшля–Теллера имеем $V_0$, так что $v = V_0/W(L) = 0$, а из формулы (11) статьи [1] имеем $\lambda = 1$. Тогда второе слагаемое в фигурных скобках (49) принимает вид

$$(2/\sqrt{\pi})[W(L)\beta]^{1/2} = (2/\sqrt{\pi})[W(L)/k_BT]^{1/2} \sim \hbar, \quad W(L) = \hbar^2\pi^2/2mL^2, \qquad (51)$$

и, следовательно, описывает влияние *квантования* энергии частицы в ящике.

Наконец, как видно из той же формулы (11) работы [1], при *малых* значениях $v = V_0/W(L)$ имеем $\lambda \approx 1 + v$; подставляя $\lambda - 1 \approx v$ во второе слагаемое в фигурных скобках (49), получим

$$(2/\sqrt{\pi})(V_0/W(L))[W(L)\beta]^{1/2} = (2/\sqrt{\pi})[\beta V_0^2/W(L)]^{1/2}, \qquad (52)$$

и, следовательно, описывает влияние *внешнего потенциала* $V_0$ на частицу. Совершенно очевидно, что оба слагаемых (51) и (52) малы по сравнению с первым слагаемым (=1) в высокотемпературном пределе $\beta \to 0$.

## Заключение

В данной статье в дополнение к результатам статьи [1] рассмотрены не только динамические, но и термодинамические свойства одномерной модели Пёшля–Теллера для сильно нелинейного квантового осциллятора, ограниченного в пространстве эффективной шириной $L$ конфайнмента. Показано, что не только на динамическом, но и термодинамическом уровне описания свойства модели Пёшля–Теллера для предельных случаев значений безразмерного параметра $\lambda = 1$ и $\lambda \to \infty$ сводятся к свойствам свободной частицы в ящике и гармонического осциллятора Блоха. Существенно, что параметр $\lambda$ определяется отношением $v = V_0/W(L)$ амплитуды потенциала $V_0$ Пёшля–Теллера к характерной энергии основного состояния $W(L) = \hbar^2\pi^2/2mL^2$ квантовой частицы массой $m$.

Показано, что при низких температурах для всех моделей (т.е. при всех значениях $\lambda$) выполняется третье начало термодинамики, а при высоких температурах в зависимости от величины $\lambda$ калорические величины – внутренняя энергия и теплоемкость воспроизводят свойства предельных моделей в соответствующих интервалах значений $\lambda$. Что касается термической величины – давления, то оно существует только для конечных значений $\lambda$ и $L$.

## Литература

## About One Interesting and Important Model in Quantum Mechanics
## II. Thermodynamic Description

Yu.G. Rudoy, E.O. Oladimedji (jr.)

*People's Friendship University of Russia (RUDN),*
*Moscow, Russian Federation, 117198, Miklukho-Maklay str., 6;*
*e-mail: rudikar@mail.ru, nockjnr@gmail.com*

In this paper the detailed investigation of one of the most interesting models in the non-relativistic quantum mechanics of one massive particle – i.e., introduced by G. Poeschl and E. Teller in 1933 – is continued; the starting point of dynamic analysis for potentials, wave functions and energy spectrum was carried on by the same authors in the paper [1]. The generalization of these results on the case of nonzero temperature includes the approximate calculation of the partition function of the model and then also the Massieu-Planck thermodynamic potential and its most important derivatives, i.e. internal energy, heat capacity and pressure. The analysis of the results is carried on separately for low- and high-temperature regions because of the lack of exact unifying expressions of the potential at all temperatures.

*Keywords*: Poeschl–Teller quantum oscillator, thermodynamic potential, thermodynamic equation of state, low- and high-temperature expansions.